\def \cm{~\rm{cm}}
\def \s{~\rm{s}}
\def \sr{~\rm{sr}}
\def \km{~\rm{km}}

\def \erg{~\rm{erg}}

\def \yr{~\rm{yr}}
\def \pc{~\rm{pc}}
\def \kpc{~\rm{kpc}}

\documentclass[]{emulateapj}

\begin{document}

\title{OBSERVED PLANETARY NEBULAE AS DESCENDANTS
OF INTERACTING BINARY SYSTEMS}
\author{Noam Soker}
%\altaffilmark{}
\altaffiltext{1} {Department of Physics, Technion$-$Israel
Institute of Technology, Haifa 32000 Israel;
soker@physics.technion.ac.il.}

\begin{abstract}
We examine recent studies on the formation rate of planetary nebulae
and find this rate to be about one-third of the formation rate of
white dwarfs. This implies than only about one-third of all
planetary nebulae that evolve to form white dwarfs are actually
bright enough to be observed. This finding corresponds with
the claim that it is necessary for a binary companion to interact
with the asymptotic giant branch stellar progenitor for the
descendant planetary nebulae to be  bright enough to be detected.
The finding about the formation rate also strengthens De Marco's
conjecture that the majority of observed planetary nebulae harbor
binary systems.   In other words, single stars almost
never form observed planetary nebulae.
\end{abstract}


% ====================================================================
\section{INTRODUCTION}
% ====================================================================
Planetary nebulae (PNs) are ionized clouds of gas outflowing from their
progenitor central star. This expanding nebula must be bright enough to
be observed in the visible band to be classified as a PN.
%%  The central star evolves to become a white dwarf (WD), and it is
%%  sufficiently hot and luminous to ionize the outflowing gas.
%%  The central star expelled the outflowing gas when it was in the
%%  final stages of its asymptotic giant branch (AGB) phase.
%%  The outer regions of PNs consist of more or less
%%  spherically symmetric outflowing gas,  carrying mass at rates of
%%  $\sim 10^{-7}-10^{-5} M_\odot \yr ^{-1}$.
Most PNs possess a global axisymmetrical structure (by axisymmetrical we refer
also to point-symmetric and multi-polar geometries), rather than a spherical
structure, in their inner region (Zuckerman \& Aller 1986; Balick 1987; Chu et al. 1987;
see Balick \& Frank 2002 and Sahai 2004 for more references).
The transition from blowing spherical wind to axi-symmetrical one by
the progenitors of PNs is most evident in PNs having an inner bright axi-symmetrical
region with a faint spherical halos around the inner region;
the inner region surface brightness is typically more than an order of magnitude
higher than that of the spherical outer region (e.g., Plait \& Soker 1990;
Bryce et al. 1994; Balick et al. 1992; Hajian et al. 1997; Corradi et al. 2003, 2004).

Frank et al. (1994) examined the possibility that PNs are formed from a
high mass loss rate episode (super-wind, or final intensive wind) caused
by a thermal helium flash. This, however, does not account for the observation
that the transition to a final intensive wind correlates with the
transition to a non-spherical mass loss geometry (Soker 2002).
The thirty-years old binary model for the formation of non-spherical PNs
(e.g., Bond et al. 1978; Fabian \& Hansen 1979; Livio et al. 1979; Morris 1981;
for review see Balick \& Frank 2002),
can account for this correlation (Soker 2004), where the binary can be a
stellar or a massive substellar object.

Over the years it has become clear that binary interaction plays a major role
in the formation and shaping of many (Bond \& Livio 1990; Iben \& Livio 1993),
or even most (Bond 2000), PNs.
Several recent studies discuss the possibility that binary systems
are necessary for the formation, mainly via the common envelope evolution,
of observed PNs
(De Marco et al. 2004; Afsar \& Bond 2005; De Marco \& Moe 2005).
De Marco \& Moe (2005) and Moe \& De Marco (2006, hereafter MD06)
reached this conclusion based on a population synthesis study and
observations hinting that most PNs have central binary systems
(Sorensen \& Pollaco 2004; De Marco et al. 2004;
Hillwig 2004; Afsar \& Bond 2005).

Soker \& Subag (2005, hereafter SS05) argued that relative to non-spherical PNs
spherical PNs are about an order of magnitude less
likely to be detected.
They argued that the majority of spherical PNs are much too faint to be
detected, and they term this population a hidden PN population.
SS05 based their claim on two arguments.
First, they noticed that all PN are expected to have a large halo
around the bright shell; the halo is the remnant of the early AGB wind.
However, only a small fraction of PNs posses such a {\it detectable} halo.
Noting the structure similarity of halos around non-spherical PNs
to that of observed spherical PNs, they then assumed that most unobserved
spherical PNs are also similar in structure to the spherical halos
around non-spherical PNs. From the work of
Corradi et al. (2003) concerning halos in PNs, SS05
conservatively deduced that the detection fraction of halos is
about an order of magnitude less than that of the bright PN shell.
Considering that the detection of halos follows the detection of the
inner bright non-spherical shell, rather than being random, the
true detection probability of halos is much below $10 \%$ of the
detection probability of the non-spherical bright shells.
Secondly, SS05 built a toy model for the luminosity evolution of
PNs, and showed that the claimed detection fraction of spherical PNs
based on halos around non-spherical PNs is compatible with
observational sensitivities. Their toy model is conservative and
overestimates the detection probability of spherical PNs. The toy
model was based on the assumption that PNs are detected by their
luminosity. However, most PNs are detected by their surface
brightness. This will make spherical PNs with their extended halo
even more difficult to detect and the expected hidden population
larger.
The recent detection of a spherical PN at a distance of only $\sim 0.5
\kpc$ (Pierce et al. 2004;  Frew \& Parker 2006) shows that faint spherical PNs can
indeed escape detection.

An interacting binary companion has two effects on the mass loss process
that cause the descendant PN to be much brighter.
(1) The companion increases the mass loss rate on the AGB. This leads to a denser
nebula, therefore a much brighter nebula.
(2) The enhanced mass loss rate continues into the post-AGB phase, making
the AGB to PN transition phase much shorter.
This effect is very important because in commonly accepted mass loss mechanisms
for AGB stars, which are based on radiation pressure on dust,
the mass loss rate is very sensitive to the stellar surface (effective)
temperature $T_{\rm eff}$
(e.g., Wachter et al.\ 2002; van Loon et al. 2005).
The star starts to leave the AGB, shrinks, and heats up when its
envelope mass decreases to $M_{\rm env} \simeq 0.1 M_\odot$.
With a mass loss rate $<10^{-5} M_\odot \yr^{-1}$ the transition from the AGB
to the PN phase, the pre-PN evolution time, will last more than $10,000\yr$,
contrary to observations.
Soker \& Harpaz (2002) and Sch\"onberner et al. (2005) previously noted
that commonly used theoretical dust-driven mass-loss models in single stars
cannot be applied during the star's departure from the AGB.
The surface brightness $S$ depends strongly on the PN radius
$S \sim R_{\rm PN}^{-3}$ (Shaw et al. 2001; Stanghellini et al. 2002, 2003;
Frew \& Parker 2006).
Therefore, a long pre-PN (post-AGB) evolutionary time, such as expected for single
AGB stars, implies a larger PN with a much lower surface brightness.

The goal of this paper is to examine whether it is indeed possible
that the majority (Bond 2000), or even all (De Marco et al. 2004; De Marco \& Moe 2005;
MD06), observed non-spherical PNs are shaped by strongly
interacting stellar (or in the minority of the cases with massive substellar)
companions, and to update the percentage of different
evolutionary routes to produce non-spherical PNs (SS05).

% ====================================================================
\section{RE-EXAMINING ESTIMATES OF PLANETARY NEBULAE FORMATION RATE}
% ====================================================================

The PN formation rate per unit volume in the solar vicinity is hard
to calculate, and published values differ substantially in the
range $\chi(PN)=0.4-8 \times 10^{-12} \pc^{-3} \yr^{-1}$
(Pottasch 1996).
We focus on the a more recent study by Phillips (2002).
Phillips (2002) finds the PN formation rate per unit volume to be
$\chi_P(PN)=2.1\times 10^{-12} \pc^{-3} \yr^{-1}$.
We find this estimate to be too high as we now explain.

(1) {\it PN Size. }
In Table 1, Phillips (2002) lists 66 close-by PNs.
Of these, 24 are in the list of PNs interacting with the ISM
compiled by Tweedy \& Kwitter (1996); 20 of these 24 are PNs strongly
interacting with the ISM.
According to Phillips, the average radius of PNs listed in
Phillips' Table 1 is $<R_P> \simeq 0.54 \pc$.
However, the average radius (along the long axis) of the 24 PNs
listed also in Tweedy \& Kwitter is $<R_{TK}> \simeq 1.45 \pc$.
This hints that on the average PN sizes and distances might be larger than
the values obtained by Phillips (2002).

(2) {\it The PN Visibility (life) time.}
We consider four factors:
(2.1) Phillips (2002) takes the PN visibility age to be equal to the
average PN radius divided by the nebular expansion velocity.
He finds $<R_P> \simeq  0.54 \pc$, and divides by
an average expansion velocity $25 \km \s^{-1}$ to obtain
the PN visibility time of $\tau_{PN}\simeq 21,000 \yr$.
However, the average age of a population will be equal to half its
life span.
(2.2) As claimed in the preceding point (1), the actual average PN radii
is larger than that found by Phillips.
(2.3) MD06 estimate the PN visibility time as a function of
metallicity and progenitor mass, and find the average value to be
$\tau_{PN} \sim 40,000 \yr$.
For comparison, Buzzoni et al. (2006) take in their population
synthesis study $\tau_{PN} \sim 30,000 \yr$.
(2.4) Because the ISM compresses the nebular gas on one side,
such PNs stay bright for a longer time than freely expanding nebulae.
About one-third of the PNs in Phillips' Table 1 show signs of interaction
with the ISM.
Considering these four factors, we estimate $\tau_{PN} \simeq 40,000 \yr$
as a more typical PN visibility (life) time.

(3) {\it Distances of close by PNs.} Since Phillips (2002) published his results,
there have been new estimates of distances to PNs in the literature.
Phillips' Table 1 has eight PNs with new estimated distances:
NGC 7009 (Fernandez et al. 2004; Sabbadin et al. 2004),
NGC 6853 (Benedict 2003),
NGC 7027 (Bains et al. 2003; Mellema 2004),
NGC 6543, NGC 7662, NGC 3242 (Mellema 2004),
NGC 3132 (Schwarz \& Monteiro 2006) and
IC 2448 (Mellema 2004; Palen et al. 2002).
On average, these distances are $\sim 1.5$  % 1.47
larger than those listed by Phillips.
In recent papers Phillips (2005a,b) updated several distances.
In Phillips (2005b) distance to four PNs are larger by an average factor
of 1.9 than those given in Phillips (2002).
In Phillips (2005a) there are distances to 29 PNs in common with Table 1
in Phillips (2002). Of these, one PN is listed with a smaller distance,
while 11 are listed with larger distances, and 17 with the same distances as those
in Phillips (2002). The average distance of these 29 PNs in Phillips (2005a)
is 1.3 times larger than the average distance of the same 29 PNs listed
in Phillips (2002).
There is no overlap between the 8+4+11=23 PNs with new and larger distances,
out of 66 PNs in Table 1 of Phillips (2002).
MD06 found their PN birth rate to be $\sim 10$ times smaller than that
deduced by Phillips and implicitly stated that the Phillips'
distance scale is too small by a factor of $\sim 2$ (De Marco, O.,
private communication, 2006).
We take our new average distance estimate to PNs to be $1.5$ times larger
than distances given by Phillips (2005).
This implies a local density of $\sim 0.3$ compared
with that obtained by Phillips.
Using the same steps as Phillips (2002) used, we find that the total
number of observed PNs in the galaxy is $\sim 9,000$ instead of $\sim 30,000$.
This is identical to the number deduced by Peimbert (1993).
Considering $\tau_{PN} \simeq 40,000 \yr$, which is $\sim 2$ times the
value used by Phillips, yields a local PN formation
rate $\chi(PN)$ which is $\sim 0.15$ times that obtained by Phillips.

(4) {\it Distances of far PNs.}
In Table 3, Phillips lists PNs at an average larger distance from
the Sun compared with those in Table 1. From these,
eleven PNs have a new distance estimate:
NGC 6578, NGC 6884 (Mellema 2004; Palen et al. 2002),
NGC 6741 (Sabbadin et al. 2005),
Vy 2-2, BD +30$^\circ$3639  (Mellema 2004),
A 20, A 15 (Emprechtinger et al. 2005),
Mz1 (Monteiro et al. 2005),
Mz3 (Smith 2003),
M 2-43 (Guzman et al. 2006), and
NGC 6302 (Meaburn et al. 2005).
The new distances are on average $\sim 2.8$ times those listed by Phillips.
This implies a PN density lower by an order of magnitude than that obtained by
Phillips (2002).
This might further support our claim that Phillips overestimated the observed-PN
formation rate by an order of magnitude.

Our new estimate of local PN formation rate is
$0.1-0.2$ times that given by Phillips, namely
$\chi_n(PN)=0.2-0.4 \times 10^{-12} \pc^{-3} \yr^{-1}$.
This should be compared with the formation rate of white dwarfs.
Liebert et al. (2005) find the WD formation rate to be
$\chi_L(WD) \sim 10^{-12} \pc^{-3} \yr^{-1}$.
$\sim 10 \%$ of these are low mass WDs, whose
progenitors did not evolve through the AGB.
On the other hand, the number of WDs can be somewhat larger
than that found by Liebert et al. (Harris et al.\ 2006).
We note that MD06 find the expected local
formation rate of PNs (hidden and observed) to be
$\sim 0.9 \times 10^{-12} \pc^{-3} \yr^{-1}$.
We conclude that our estimate of the {\it observed} PN formation rate
is $\sim 0.3$ times the formation rate of WD by AGB stars, or of all PNs.
The rest $\sim 70 \%$ of WDs formed by AGB stars are descendants
of hidden PNs.

% ====================================================================
\section{CONSTRAINING THE PROPERTIES OF HIDDEN PLANETARY NEBULAE}
% ====================================================================

 The discussion in \S1 and \S2, and the papers cited there,
strongly suggest that there is indeed a {\it hidden PN population},
defined as PNs that originate from AGB stars and are now ionized
by the same central stars, but which are much too faint to be detected.
They are formed from stars that did not go through a final intensive mass
loss rate episode at the end of their AGB phase, most likely because they did
not interact with a stellar companion or a brown dwarf or a massive planet.

Limitations from different surveys can be used to constrain the properties
of the hidden-PN progenitors.
Surveys for PNs in the $H\alpha$ emission line that have high enough angular
resolution are sensitive down to a surface brightness of 2 Rayleighs, which equals
$S_s \sim 5 \times 10 ^{-7} \erg \cm^{-2} \s^{-1} \sr^{-1} $ (Parker et al. 2005).

We consider a spherical PN formed by an AGB progenitor which had a slowly
varying mass loss rate, which decreased by a large factor as it
terminated the AGB.
The ionization of the nebula by the central star, i.e., the PN phase, starts
a time $\tau_p$ after the termination of the AGB wind.
The nebula, i.e., the AGB wind, is characterized by a mass loss rate $\dot M_h$,
a speed $v$ and density $\rho(r)= \dot M_h/4 \pi r^2 v$.
Because the pre-PN time period is long, and as a consequence the internal
low density cavity is large, we neglect the compression of the slow wind by
the fast stellar wind blown during the PN phase.
This could increase the surface brightness to some extent.
The surface brightness has a maximum at a projected distance $R=v \tau_p$.
The surface brightness is given by integrating the emissivity, in
units of $\erg \s^{-1} \cm^{-3}$, along the line of sight through the nebula
\begin{equation}
S_{PN}= \frac{1}{4 \pi} \int_{-\infty}^{\infty} n_p n_e \alpha_{H_\alpha} d l,
\label{int1}
\end{equation}
where $n_p$ is the proton density, $n_e$ is the electron density,
and $\alpha_{H_\alpha}$ is the line recombination coefficient (Case B)
for $H\alpha$.
After performing a standard integration along the line of sight, the
surface brightness in $H\alpha$ is found to be
\begin{eqnarray}
S_{PN}= 7.5 \times 10^{-6} % 7.46
\left( \frac {\dot M_h}{10 ^{-6} M_\odot \yr^{-1}} \right)^2
\left( \frac {\tau_p}{10^4 \yr} \right)^{-3} \nonumber \\
\times
\left( \frac {v}{10 \km \s^{-1}} \right)^{-5}
\erg \cm^{-2} \s^{-1} \sr^{-1}. \qquad
\label{sb1}
\end{eqnarray}

To avoid detection most of these PNs should have a surface brightness of
$S_{PN} < S_s \sim 5 \times 10 ^{-7} \erg \cm^{-2} \s^{-1} \sr^{-1} $.
The constraint on the pre-PN (post-AGB) evolution time period (the transition
from the AGB to the PN phase) reads
\begin{equation}
\tau_p \ga 25,000
\left( \frac {\dot M_h}{10 ^{-6} M_\odot \yr^{-1}} \right)^{2/3}
\left( \frac {v}{10 \km \s^{-1}} \right)^{-5/3} \yr
\label{taup}
\end{equation}

Theories of mass loss driven by radiation pressure on dust predict
very strong dependence on the effective temperature of the star.
Wachter et al. (2002) find $\dot M \sim T_{\rm eff}^{-7}$
for carbon-rich AGB stars.
van Loon et al. (2005) empirically find the dependence of mass loss rate
of oxygen-rich AGB stars on effective temperature to be
$\dot M \sim T_{\rm eff}^{-6.3}$.
AGB stars start to contract when their envelope mass is $\sim 0.1 M_\odot$.
This implies that for most single AGB stars, the mass loss rate
decreases to $\sim 10^{-6} M_\odot \yr^{-1}$ when the envelope mass is
several$\times 0.01 M_\odot$.
Therefore, a pre-PN evolution time of several$\times 10^4 \yr$ is expected
for single stars, and this is what Wachter et al. (2002)
find for an evolution of a carbon-rich
star having a main sequence mass of $1.3 M_\odot$.
Oxygen-rich stars will have lower mass loss rate,
and evolve over a longer time period.
This shows that the condition for PNs formed by single stars not
to be detected by present surveys (eq. \ref{taup}) is fulfilled by
most, or even all, single star progenitors, but not to a large extent,
implying that deeper observations will reveal more faint PNs,
most of them spherical.

% ====================================================================
\section{DISCUSSION AND SUMMARY}
% ====================================================================

In proposing that the majority of PNs are descendants of binary systems
(Bond 2000), the primary stimulus of De Marco et al. (2004), De Marco \& Moe (2005)
and MD06 derives from observations that suggest a large
fraction of PNs have a central binary stellar system.
The main motivation for our study and finding that it is
indeed possible that most {\it observed} PNs are descendants of binary
systems is the firmly based theoretical result that single stars cannot
form non-spherical PNs (Soker 2006 and references therein).
Whereas $>95 \%$ of observed PNs are non-spherical, population synthesis
studies indicate that $\sim 25 \%$ of all AGB stars go through a common
envelope interaction (Yungelson et al. 1993; Han et al.\ 1995), and
an additional $\sim 15 \%$ of AGB stars interact with a close stellar
companion that avoids a common envelope (Soker \& Rappaport 2000).
In these two processes the stellar companion not only shapes the
AGB wind but also increases the mass loss rate and shortens the
pre-PN evolution time.
Both effects lead to a much brighter descendant PN.
Shaping by wide companions or planets are not expected to increase the mass
loss rate from AGB stars by much. Also, it is not clear that there are enough
massive planets and brown dwarfs to shape $\sim 15 \%$ of all PNs as suggested
by SS06.

Re-examining the PN formation rate derived by Philips (2002), we found
that Phillips overestimated the PN formation rate by a factor of $\sim 7$.
Our results indicate the formation rate of PNs to be $\sim 1/3$
of the WD formation rate (Liebert et al. 2005).
If this result holds, then we must conclude that the hidden PN population
comprise $\sim 60-70 \%$ of all PNs. These are mostly, but not solely,
spherical PNs.
We also conclude that the vast majority of observed PNs are descendants of
strongly interacting binary systems, but these
comprise only $\sim 1/3$ of all PNs.
Updating the results of SS05, we conclude that among the observed PNs,
only $\sim 3 \%$ are spherical; these are descendants of single star evolution.
The rest, $\sim 97 \%$ of observed PNs, are non-spherical,
with $\sim 60-70 \%$ being descendants of common envelope evolution, and
$\sim 30-40 \%$ being descendant of strongly interacting binary systems
that avoided common envelope evolution.
Many post-AGB stars, which most likely are interacting binary stars
that avoided a common envelope, have circumstellar gas and dust concentrated
near their equatorial plane (De Ruyter et al. 2006). These systems show that
many post-AGB stars with a stellar companion that avoided  a common envelope
are capable of causing non-spherical mass loss geometry.

Our results strengthen the conjecture of De Marco and collaborators
(De Marco et al. 2004; De Marco \& Moe 2005; MD06) that most
{\it observed} PNs are formed by stellar binary systems
(in a minority of the cases the companion can be a brown dwarf
of a massive planet).
That is, single stars rarely form observed PNs.
We join SS05 in predicting that {\it very} deep observations
will detect more faint PNs, most of which are spherical.
We also predict that stars similar to those of the central stars of PNs
but with no bright nebula around them will be detected.

% ====================================================================
\acknowledgments
This research was supported in part by the Asher
Fund for Space Research at the Technion.
% ====================================================================

\end{document}